\documentclass[12pt, a4paper]{article}
\usepackage{a4wide,amssymb,amsmath,amscd,}

\begin{document}

\topmargin -2pt

\baselineskip 20pt

\begin{flushright}
{\tt quant-ph/0308156} \\
\end{flushright}

\vspace{1mm}

\begin{center}

{\Large \bf  Quantum Entanglement under Lorentz Boost }\\

\vspace{9mm}

{\sc Daeho Lee~ and~ Ee Chang-Young}\footnote{cylee@sejong.ac.kr}\\
{\it Department of Physics, Sejong University, Seoul 143-747, Korea}\\

\vspace{18mm}

{\bf ABSTRACT}
\end{center}
In order to understand the characteristics of quantum entanglement
of massive particles under Lorentz boost, we first introduce a
relevant relativistic spin observable, and evaluate its
expectation values for the Bell states under Lorentz boost. Then
we show that maximal violation of the Bell's inequality can be
achieved by properly adjusting the directions of the spin
measurement even in a relativistically moving inertial
frame.
 Based on this we infer that the entanglement information is preserved
 under Lorentz boost as a form
of correlation information determined by the transformation
characteristic of the Bell state in use.
\\

\vfill


\thispagestyle{empty}

\newpage

\section*{I. Introduction}

Quantum entanglement is a novel feature of quantum physics when
compared with classical physics. It demonstrates the nonlocal
character of quantum mechanics and is the very basis of quantum
information processing such as quantum computation and quantum
cryptography. Up until recently quantum entanglement was
considered only within the non-relativistic regime. Then, starting
with the work of \cite{czachor} there have been quite a lot of
works investigating the effect of quantum entanglement measured in
an inertial frame moving with relativistic speed
\cite{am,pst,ging,tu,ahn,rsw,crsw,terno,czw,ps,czachor2,peres,bga,ahnk}.

Consider two spin half particles with total spin zero moving in
opposite directions. Suppose the spin component of each particle
is measured in the same direction by two observers in the
laboratory(lab) frame. Then the two spin components have opposite
values in whichever direction the spin measurements are performed.
This is known as the EPR(Einstein, Podolsky, Rosen) correlation
and is due to the isotropy of the spin singlet state. Is the EPR
correlation valid even for the two observers sitting in a moving
frame which is Lorentz boosted relativistically with respect to
the lab frame? This issue has been investigated from various
aspects by many people including the above quoted authors.
However, the answer for this question has not been clarified so
far.

In \cite{czachor}, Czachor  considered the spin singlet of two
spin-$\frac{1}{2}$ massive particles moving in the same direction.
He introduced the concept of relativistic spin observable which is
closely related with the spatial components of the Pauli-Lubanski
vector. For two observers in the lab frame measuring the spin
component of each particle in the same direction, the expectation
value of the joint spin measurement, i.e., the expectation value
of the tensor product of relativistic spin observable of each
constituent particles, became dependent on the boost velocity.
Only when the boost speed reaches that of light, or when the
direction of the spin measurements is perpendicular to the boost
direction, the expectation value becomes $-1$. Thus only in these
limiting cases the results seem to agree with the EPR correlation.
%
%
%
 Czachor considered only the changes in the spin operator part
by defining a new relativistic spin operator. There, the state
does not need to be transformed since the observer is at rest.
%
%
%

%
Starting a year ago there appeared a flurry of papers
investigating the effect of the Lorentz boost or the Wigner
rotation on entanglement. Here, we mention some of them that are
directly related to the issue in our work.

Alsing and Milburn \cite{am} considered the entanglement of two
particles moving in opposite directions and showed that the Wigner
rotation under Lorentz boost is a local unitary operation, with
which Dirac spinors representing the two particles transform. And
they reached a conclusion that the entanglement is Lorentz
invariant due to this unitary operation.

Gingrich and Adami \cite{ging} investigated the entanglement
between the spin and momentum parts of two entangled particles.
They concluded that the entanglement of the spin part is carried
over to the entanglement of the momentum part under Lorentz boost,
although the entanglement of the whole system is Lorentz invariant
due to unitarity of the transformation.
However, the concept of the reduced density matrices with
traced-out momenta used in that work drew some criticism recently
\cite{czw}.
%
%
%
%

%
Terashima and Ueda \cite{tu} considered the effect of Wigner
rotation on the spin singlet and evaluated the Bell observable
under Lorentz boost. They concluded that although the degree of
the violation of the Bell's inequality is decreased under Lorentz
boost, the maximal violation of the Bell's inequality can be
obtained by properly adjusting the directions of the spin
measurements in the moving frame. They also claimed that the
perfect anti-correlation of the spin singlet seen in the EPR
correlation is maintained for appropriately chosen spin
measurements directions depending on the Lorentz boost, even
though the EPR correlation is not maintained when the directions
of spin measurements remain the same.
%

%
%
In \cite{tu}, Terashima and Ueda considered the changes in the
states only. Their spin operator has the same form as the
non-relativistic spin operator. In this sense their result that
the maximal violation of Bell's inequality can be achieved even in
the moving frame was somewhat expected due to the unitarity of the
state transformation. In fact, \cite{am} and \cite{tu} considered
the changes in the states only, and the both reached a similar
conclusion that the entanglement can be preserved under Lorentz
boost.

However, if one considers the changes of the spin operator under
Lorentz transformation as Czachor did in \cite{czachor}, the story
becomes different: The Bell's inequality might not be violated.
I.e., the entanglement may not be preserved under the
transformation.
%
%
%
%

%
In a general situation, one has to consider the both, the changes
in the spin operator and the changes in the states. This was done
by Ahn etal. in \cite{ahn}. Ahn {\it et al.} \cite{ahn} calculated
the Bell observable for the Bell states under Lorentz boost, and
showed that the Bell's inequality is not violated in the
relativistic limit. They used the Czachor's relativistic spin
operator and transformed the state under Lorentz boost
accordingly. Their result strongly suggested that the entanglement
is not preserved under Lorentz boost.
They further concluded
\cite{ahnk} that quantum entanglement is not invariant under
Lorentz boost based on the evaluation of the entanglement fidelity
\cite{schum}.

Here, we would like to note that the spin operator used in
\cite{ahn} is not as general as it should be. This is because the
spin operator used in \cite{ahn} is the same one as the Czachor's
\cite{czachor} which is a spin operator for a restricted
situation, that we call the Czachor's limit in this paper.

%
%
%
%
%
%

%
In this paper, we consider the changes in the spin operator under
the Lorentz boost in a general situation compared with \cite{ahn}.
We first formulate the relativistic spin observable based on the
sameness of the expectation values of one-particle spin
measurement evaluated in two relative reference frames, one in the
lab frame in which the particle has a velocity $\vec{v}$ and the
observer is at rest, the other in the moving frame Lorentz boosted
with $\vec{v}$, in which the particle is at rest and the observer
is moving with a velocity $-\vec{v}$. Applying the relativistic
spin observable for the two-particle spin singlet state we
evaluate the expectation value of the joint spin measurement. Then
we calculate the values of the Bell observable for the Bell
states. The values of the Bell observable decreased as the boost
velocity becomes relativistic. However, we find a new set of spin
measurement axes with which the Bell's inequality is maximally
violated. This seems to imply that the information on the
correlation due to entanglement is kept even in the moving frame.
In fact, under a Lorentz boost certain entangled states transform
into combinations of different entangled states. However, in
certain directions of spin measurements the above combinations of
states become eigenstates of these spin operators. In this manner
the correlation information in one frame is maintained in other
frames.

The paper organizes as follows. In section II, we formulate the
relativistic spin observable. Then in section III, we evaluate the
expectation value of the joint spin measurement for spin singlet.
In section IV, we find a new set of spin measurement axes for a
spin singlet state, with which the Bell's inequality is maximally
violated even under Lorentz boost. In section V, we show that the
same thing can be done for the other Bell states. We conclude with
discussion in section VI.
\\

%
%
%
%
%
%
%
%
%
%
%
%

\section*{II. Relativistic spin observable}

In this section, we consider a spin measurement of a massive
particle viewed from two different inertial reference frames: one
in the lab frame where the particle has a certain velocity, the
other in the moving frame where the particle is at rest. In order
to make the particle at rest, the moving frame is Lorentz boosted
in the opposite direction of the particle's velocity in the lab
frame just to compensate the particle's motion in the lab frame.

Since we are just considering the same measurement in the two
inertial frames, the respective expectation values of this
measurement observed in the two frames should be the same.
%
%
\begin{equation}
\left<\Psi_{\vec{p}}\left| \frac{\vec{a}\cdot
\vec{\sigma_{p}}}{|\lambda(\vec{a}\cdot \vec{\sigma_{p}}) |}
\right|\Psi_{\vec{p}}\right>_{\text{lab}}
=\left<\Psi_{\vec{p}=0}\left| \frac{\vec{a_{p}}\cdot
\vec{\sigma}}{|\lambda(\vec{a_{p}}\cdot \vec{\sigma}) |}
\right|\Psi_{\vec{p}=0}\right>_{\text{rest}} \label{ce1}
\end{equation}
 Here, $\vec{a}$ and $\vec{p}$ are the spin measurement axis
 and the momentum of the particle respectively in the lab frame,
 and
 $\vec{a_{p}}$ is inversely Lorentz boosted vector of $\vec{a}$
 by $-\vec{p}$.
$|\Psi_{\vec{p}}>$ and $|\Psi_{\vec{p}=0}>$ are the wave functions
of the particle in the lab and moving frames, respectively.

Now, the two wave functions are related by
\begin{equation}
|\Psi_{\vec{p}}> =U(L(\vec{p}))~ |\Psi_{\vec{p}=0}>
=U(R(\hat{p}))~U(L_{z}(|\vec{p}|))~ |\Psi_{\vec{p}=0}>,
\label{ce2}
\end{equation}
since for an arbitrary four momentum $p$ it can be written as
$p=R(\hat{p})L_{z}(|\vec{p}|)k$ where $k=(m,0,0,0)$ is the four
momentum of the particle at rest, $U(L(\vec{p}))$ and
$L_{z}(|\vec{p}|)$ are the Lorentz boosts along $\vec{p}$ and the
$z$-axis respectively, and $R(\hat{p})$ is a rotation of $\hat{z}$
to $\hat{p}$.

%
If we decompose the particle's wave functions into their spatial
and spin parts, for instance $|\Psi_{p=0}>=|0>\otimes
|\chi_{p=0}>,$ using the relation (\ref{ce1}) we can obtain the
relation between the relativistic spin operator $\vec{\sigma_{p}}$
in the lab frame and the non-relativistic spin operator
$\vec{\sigma}$ in the particle at rest frame (the moving frame) in
terms of the spin measurement axis $\vec{a}$ in the lab frame and
its Lorentz transformed spin measurement axis $\vec{a_{p}}$ in the
moving frame.
%
%
\begin{eqnarray}
& & \left<\chi_{\vec{p}=0}\left| \otimes \left< \vec{p} =  0\left|
\frac{\vec{a_{p}}\cdot \vec{\sigma}}{|\lambda(\vec{a_{p}}\cdot
\vec{\sigma}) |}\right|\vec{p}
=  0\right>\otimes\right|\chi_{\vec{p}=0}\right> _{\text{rest}}  \nonumber\\
& & = \left<\Psi_{\vec{p}=0}\left|
U^{\dagger}(L_{z}(|\vec{p}|))U^{\dagger}(R(\hat{p}))
 \left[\frac{\vec{a}\cdot \vec{\sigma_{p}}}{|\lambda(\vec{a}\cdot
\vec{\sigma_{p}}) |}\right] U(R(\hat{p}))~U(L_{z}(|\vec{p}|))
\right|\Psi_{\vec{p}=0}\right>_{\text{lab}}
\label{ce3}
\end{eqnarray}
Also, $U(L_{z}(|\vec{p}|))~|\Psi_{p}=0>=|p_{z}> \otimes~
|\chi_{p_{z}}>$ thus we have
%
%
\begin{eqnarray}
 & & <\vec{p}=0|\vec{p}=0>\left<\chi_{\vec{p}=0}\left|
\frac{\vec{a_{p}}\cdot \vec{\sigma}}{|\lambda(\vec{a_{p}}\cdot
\vec{\sigma}) |}\right|\chi_{\vec{p}=0}\right>
_{\text{rest}}  \nonumber\\
 & &  =\left<\chi_{p_{z}}\left|\otimes \left
<p_{z}\left| U^{\dagger}(R(\hat{p})) \left[\frac{\vec{a}\cdot
\vec{\sigma_{p}}}{|\lambda(\vec{a}\cdot \vec{\sigma_{p}})
|}\right] U(R(\hat{p})) \right|p_{z} \right >\otimes \right|
\chi_{p_{z}}\right>_{\text{lab}} \nonumber\\
& & =\left<\chi_{p_{z}}\left|U^{\dagger}(R(\hat{p}))\otimes \left
<\vec{p}\left| \left[\frac{\vec{a}\cdot
\vec{\sigma_{p}}}{|\lambda(\vec{a}\cdot \vec{\sigma_{p}})
|}\right] \right|\vec{p} \right >\otimes U(R(\hat{p})) \right|
\chi_{p_{z}}\right>_{\text{lab}}
\nonumber\\
& &
=<\vec{p}~|~\vec{p}>\left<\chi_{p_{z}}\left|U^{\dagger}(R(\hat{p}))
\left[\frac{\vec{a}\cdot \vec{\sigma_{p}}}{|\lambda(\vec{a}\cdot
\vec{\sigma_{p}}) |}\right] U(R(\hat{p})) \right|
\chi_{p_{z}}\right>_{\text{lab}} \label{ce4}
\end{eqnarray}
Here, we notice that the spin wave function in the particle at
rest frame is not affected under arbitrary Lorentz boost
$L(\vec{p})$ since the Wigner angle due to the Lorentz boost in
this case is zero:
%
%
\begin{equation}
|\Psi_{\vec{p}}>=U(L(\vec{p}))|\Psi_{\vec{p}=0}>
= U(L(\vec{p})) |0> \otimes \left(%
\begin{array}{c}
\alpha \\
 \beta \\
\end{array}%
\right)
=|\vec{p}> \otimes \left(%
\begin{array}{c}
\alpha \\
 \beta \\
\end{array}%
\right).
\end{equation}
Namely, $ | \chi_{\vec{p}=0}> _{\text{rest}} = |
  \chi_{p_{z}}>_{\text{lab}} , $ and thus from (\ref{ce3}) and
(\ref{ce4}),  we obtain the following relation.
%
%
\begin{equation}
U^{\dagger}(R(\hat{p})) \left[\frac{\vec{a}\cdot
\vec{\sigma_{p}}}{|\lambda(\vec{a}\cdot \vec{\sigma_{p}})
|}\right] U(R(\hat{p})) =\frac{\vec{a_{p}}\cdot
\vec{\sigma}}{|\lambda(\vec{a_{p}}\cdot \vec{\sigma}) |}
\frac{<\vec{p}=0|\vec{p}=0>}{<\vec{p}~|~\vec{p}>}. \label{ce6}
\end{equation}

Based on the above observation, we define the relativistic spin
observable as
%
%
\begin{equation}
\hat{a} \equiv \vec{a}\cdot \vec{\sigma_{p}}/|\lambda(\vec{a}\cdot
\vec{\sigma_{p}})| \equiv  U(R(\hat{p})) \frac{\vec{a_{p}}\cdot
\vec{\sigma}}{|\lambda(\vec{a_{p}}\cdot \vec{\sigma})|}
U^{\dagger}(R(\hat{p})). \label{ce7}
\end{equation}
Here, $R(\hat{p})$ is the rotation from the $z$-axis to the
direction of $\vec{p}$, which can be written as
%
%
\begin{equation}
R(\hat{p}) =R_{z}(\phi_{p})R_{y}(\theta_{p}) =  \left(%
\begin{array}{ccc}
  \cos\phi_{p}\cos\theta_{p} & -\sin \phi_{p} & \cos\phi_{p}\sin\theta_{p} \\
  \sin \phi_{p}\cos\theta_{p} &\cos\phi_{p}  & \sin \phi_{p}\sin\theta_{p} \\
  -\sin\theta_{p} & 0 & \cos\theta_{p} \\
\end{array}%
\right),
\label{ce8}
\end{equation}
and
%
%
\begin{equation}
U(R(\hat{p}))=\exp(-i\phi_{p}\sigma_{z}/2)\exp(-i\theta_{p}\sigma_{y}/2).
\end{equation}
The spin measurement axis in the moving frame, $\vec{a}_{p}$, is
given by the spatial part of
${a}_{p}=[R(\hat{p})L_{z}(|\vec{p}|)]^{-1}a$, where
$p=L_{\hat{p}}(|\vec{p}|)k = R(\hat{p})L_{z}(|\vec{p}|)k$ with
$k=(m,0,0,0)$, and the spin measurement axis in the lab frame,
$\vec{a}$, is the spatial part of $a$.
Putting all this together, the relativistic spin observable can be
expressed as follows.
%
%
\begin{eqnarray}
\hat{a} &= &U(R(\hat{p})) \frac{\vec{a_{p}}\cdot
\vec{\sigma}}{|\lambda(\vec{a_{p}}\cdot \vec{\sigma})|}
U^{\dagger}(R(\hat{p})) \nonumber\\
& = & \frac{\vec{a_{p}}}{|\lambda(\vec{a_{p}}\cdot \vec{\sigma})|}
\cdot [\exp(-i\phi_{p}\sigma_{z}/2)\exp(-i\theta_{p}\sigma_{y}/2)~
\vec{\sigma}~\exp(i\theta_{p}\sigma_{y}/2)\exp(i\phi_{p}\sigma_{z}/2)]
\nonumber\\
& = & \vec{a_{p}}\cdot
R(\hat{p})\vec{\sigma}/|\lambda(\vec{a_{p}}\cdot\vec{\sigma} )|
\label{ce10}
\end{eqnarray}
\\

%
%
%
%
%
%
%
%
%

\section*{III. Relativistic joint spin measurement for spin singlet }

In this section, we apply the relativistic spin observable defined
in the previous section to a spin singlet state which consists of
two massive spin-$\frac{1}{2}$ particles.

 First, we consider a simple case in which the spin measuring device
 is fixed in the lab frame and
 the both particles are moving with the same
 velocity in the lab frame. This is the same set-up what Czachor
 considered in his work \cite{czachor}.
The spin measuring axes are in the direction of $\vec{a}$ for
particle 1, and in the direction of $\vec{b}$ for particle 2. We
choose the particles' moving direction as the $+z$ axis. Then the
expectation value of joint spin measurement for the particles can
be expressed as
%
%
\begin{eqnarray}
<\hat{a}\otimes \hat{b}> & = & \left<\Psi\left|\frac{\vec{a}\cdot
\sigma_{p}}{|\lambda(\vec{a}\cdot \sigma_{p})|} \otimes
\frac{\vec{b}\cdot \sigma_{p}}{|\lambda(\vec{b}\cdot \sigma_{p})|}
  \right|\Psi \right> \nonumber\\
 & =& \left<\Psi\left|  \frac{\vec{a_{p}}\cdot
R(\hat{p})\vec{\sigma}}{|\lambda(\vec{a_{p}}\cdot\vec{\sigma} )|}
\otimes \frac{\vec{b_{p}}\cdot
R(\hat{p})\vec{\sigma}}{|\lambda(\vec{b_{p}}\cdot\vec{\sigma} )|}
 \right|\Psi \right> \nonumber\\
& = & \left<\Psi\left|  \frac{\vec{a_{p}}\cdot
\vec{\sigma}}{|\lambda(\vec{a_{p}}\cdot\vec{\sigma} )|} \otimes
\frac{\vec{b_{p}}\cdot
\vec{\sigma}}{|\lambda(\vec{b_{p}}\cdot\vec{\sigma} )|}
 \right|\Psi \right>
\label{ce11}
\end{eqnarray}
where the state function is given by
$|\Psi>=\frac{1}{\sqrt{2}}~(|\vec{p},\frac{1}{2}>|\vec{p},\frac{-1}{2}>
 -|\vec{p},\frac{-1}{2}>|\vec{p},\frac{1}{2}> )$.
In the last step, we used $R(\hat{p})=1$  since $\hat{p} =
\hat{z}$ in the present case.
 The measuring axis $\vec{a_{p}}$ in the moving frame is given by the spatial part of
 Lorentz transformed $a_{p}^{~\mu}=L_{z}(- \xi )_{~\nu}^{\mu}~a^{\nu}$
 where $\tanh\xi \equiv \beta_{p}$
 representing the velocity of the particles:
%
%
\begin{equation}
a_{p}=\left(%
\begin{array}{cccc}
  \cosh\xi & 0 & 0 & -\sinh\xi \\
  0 & 1 & 0 & 0 \\
  0 & 0 & 1 & 0 \\
  -\sinh\xi & 0 & 0 & \cosh\xi \\
\end{array}%
\right)  \left(%
\begin{array}{c}
  0 \\
  a_{x} \\
  a_{y} \\
  a_{z} \\
\end{array}%
\right)= \left(%
\begin{array}{c}
  -a_{z}\sinh\xi \\
  a_{x} \\
  a_{y} \\
  a_{z}\cosh\xi \\
\end{array}%
\right).
\label{ce12}
\end{equation}
Since the magnitude of $\vec{a_{p}}$ is the same as that of the
eigenvalue of
 $\vec{a_{p}}\cdot \vec{\sigma}$,
we get
$|\lambda_{a_{p}}|=\sqrt{a_{x}^{2}+a_{y}^{2}+a_{z}^{2}\cosh^{2}\xi}
=\sqrt{1+a_{z}^{2}\sinh^{2}\xi}$.
Thus, the relativistic spin observable for particle 1 in the
present case is given by
%
%
\begin{equation}
\hat{a} \equiv  \frac{\vec{a} \cdot
\vec{\sigma_{p}}}{|\lambda(\vec{a} \cdot \vec{\sigma_{p}})|} =
\frac{\vec{a_p} \cdot \vec{\sigma}}{|\lambda(\vec{a_p} \cdot
\vec{\sigma})|}
=\frac{a_{x}\sigma_{x}+a_{y}\sigma_{y}+a_{z}\sigma_{z}\cosh\xi}
{\sqrt{1+a_{z}^{2}\sinh^{2}\xi}}.
\label{ce13}
\end{equation}
The same is for particle 2. Thus, the expectation value of the
joint spin measurement (\ref{ce11}) is given by
%
%
\begin{eqnarray}
<\hat{a}\otimes \hat{b}>  & = &
\frac{<\Psi|(a_{x}\sigma_{x}+a_{y}\sigma_{y}+a_{z}\sigma_{z}\cosh\xi)
\otimes(b_{x}\sigma_{x}+b_{y}\sigma_{y}+b_{z}\sigma_{z}\cosh\xi)
|\Psi>}{\sqrt{1+a_{z}^{2}\sinh^{2}\xi}\sqrt{1+b_{z}^{2}\sinh^{2}\xi}}\nonumber\\
& = & -\frac{(a_{x}b_{x}+a_{y}b_{y}+a_{z}b_{z}\cosh^{2}\xi)}
{\sqrt{1+a_{z}^{2}\sinh^{2}\xi}\sqrt{1+b_{z}^{2}\sinh^{2}\xi}}
\label{ce14}
\end{eqnarray}
where we used  $<\Psi|~\sigma_{i} \otimes
\sigma_{j}~|\Psi>=-\delta_{ij}$ for $i,j=x,y,z$. The (\ref{ce14})
agrees with the Czachor's result.

In order to see whether the Bell's inequality is still maximally
violated in this case, we now consider the so-called Bell
observable $C(a,a',b,b')$ defined as \cite{czachor}
%
%
\begin{equation}
C(a,a',b,b') \equiv  <\hat{a} \otimes \hat{b}>+<\hat{a} \otimes
\hat{b'}>+<\hat{a'} \otimes \hat{b}>-<\hat{a'} \otimes \hat{b'}>.
\label{bell-obs}
\end{equation}
For maximal violation, we choose the following set of vectors for
spin measurements
%
%
\begin{eqnarray}
\vec{a} &=& (0,\frac{1}{\sqrt{2}}~,~\frac{1}{\sqrt{2}})~,~~
\vec{a'}  = (0,-\frac{1}{\sqrt{2}}~,~\frac{1}{\sqrt{2}}),
\nonumber\\
\vec{b}&=& (0~,~0~,~1) ~, ~~~ \vec{b'}=(0~,~1~,~0),
\label{ce16}
\end{eqnarray}
which yields $ | C(a,a',b,b') | = 2 \sqrt{2} $ in the
non-relativistic case.
Using (\ref{ce14}), we get the Bell observable for the above
vector set as
%
%
\begin{equation}
C(a,a',b,b')
 =-\frac{2(1+\cosh\xi)}{\sqrt{2+\sinh^{2}\xi}}.
\label{ce17}
\end{equation}
We see that $|C(a,a',b,b')|$ approaches 2 in the relativistic
limit $\xi \rightarrow \infty$, thereby the Bell's inequality is
not violated in this relativistic limit.

Next, we consider a more general situation in which the two
particles of the spin singlet move in opposite directions in the
lab frame and the two observers for particle 1 and 2 are sitting
in the moving frame Lorentz boosted with respect to the lab frame
in the direction perpendicular to the particles' movements.
Here, we choose particle 1 and 2 are moving in the $+z$ and $-z$
directions respectively in the lab frame, and the moving frame in
which the two observers for particle 1 and 2, Alice and Bob, are
sitting is Lorentz boosted to the $-x$ direction.
Now, the expectation value of the joint spin measurements
performed by Alice and Bob can be expressed as
%
%
\begin{eqnarray}
<\hat{a}\otimes \hat{b}> & =& \left<\Phi\left|\frac{\vec{a}\cdot
\vec{\sigma}_{\Lambda p}}{|\lambda(\vec{a}\cdot
\vec{\sigma}_{\Lambda p})|} \otimes \frac{\vec{b}\cdot
\vec{\sigma}_{\Lambda Pp}}{|\lambda(\vec{b}\cdot
\vec{\sigma}_{\Lambda Pp})|}
  \right|\Phi \right>  \nonumber\\
 & = & \left<\Phi\left|\frac{\vec{a}_{\Lambda p}\cdot R(\vec{p}_{\Lambda})
~\vec{\sigma}}{|\lambda(\vec{a}_{\Lambda p}\cdot \vec{\sigma})|}
\otimes \frac{\vec{b}_{\Lambda Pp}\cdot R(\vec{p}_{\Lambda P})
~\vec{\sigma}}{|\lambda(\vec{b}_{\Lambda Pp}\cdot \vec{\sigma})|}
  \right|\Phi \right>.
\label{ce18}
\end{eqnarray}
Here,
 $|\Phi>=U(\Lambda)~|\Psi>$ where
 $ |\Psi> = \frac{1}{\sqrt{2}}~(|\vec{p},\frac{1}{2}>|-\vec{p},\frac{-1}{2}>
 -|\vec{p},\frac{-1}{2}>|-\vec{p},\frac{1}{2}> )$, and
 $\Lambda$ is the Lorentz boost performed to Alice and Bob(in the moving frame).

In general, the effect of a Lorentz transformation to a state can
be expressed as \cite{sw}
%
%
\begin{equation}
U(\Lambda)~|~p~,\sigma>=\sum_{\sigma'}D_{\sigma \sigma
'}(W(\Lambda , p ))~ |~\Lambda p~ ,~ \sigma'> .
\label{LB-state}
\end{equation}
The explicit form for the singlet is given by
%
%
\begin{equation}
U(\Lambda)~|\Psi>=\cos \Omega_{p} |\Psi_{\Lambda}^{(-)}> + \sin
\Omega_{p} |\Phi_{\Lambda}^{(+)}> \label{LB-singlet}
\end{equation}
where
\[
|\Psi_{\Lambda}^{(-)}>= \frac{1}{\sqrt{2}}(|\Lambda
p,\frac{1}{2}>|\Lambda Pp,\frac{-1}{2}>-|\Lambda
p,\frac{-1}{2}>|\Lambda Pp,\frac{1}{2}> ),
\]
\[
|\Phi_{\Lambda}^{(+)}>= \frac{1}{\sqrt{2}}(|\Lambda
p,\frac{1}{2}>|\Lambda Pp,\frac{1}{2}>+~|\Lambda
p,\frac{-1}{2}>|\Lambda Pp,\frac{-1}{2}> ),
\]
and $\Omega_{p}$ is the Wigner angle due to Lorentz boost $\Lambda
$ performed to a particle with momentum $\vec{p}$ and is given
explicitly by
%
%
\begin{equation} \tan \Omega_{p}=\frac{\sinh \xi \sinh
\chi}{\cosh\xi+\cosh\chi}, ~~ \text{with} ~~ \tanh \xi =  \beta_p
~, ~ \tanh \chi = \beta_{\Lambda}. \label{ce21}
\end{equation}
Here, $P$ is the space inversion operator given by
 $P=\left(%
\begin{array}{cccc}
  1 & 0 & 0 & 0 \\
  0 & -1 & 0 & 0 \\
  0 & 0 & -1 & 0 \\
  0 & 0 & 0 & -1 \\
\end{array}%
\right)$,
and thus $Pp$ is given by
$\left(%
\begin{array}{c}
  \sqrt{p^{2}+m^{2}} \\
  0 \\
  0 \\
  -p \\
\end{array}%
\right)$.
Other expressions appeared above are given by
\begin{eqnarray}
\Lambda p & = & \left(%
\begin{array}{cccc}
  \cosh\chi & \sinh\chi & 0 & 0 \\
  \sinh\chi & \cosh\chi & 0 & 0 \\
  0 & 0 & 1 & 0\\
  0 & 0 & 0 & 1 \\
\end{array}%
\right)\left(%
\begin{array}{c}
  \sqrt{p^{2}+m^{2}} \\
  0 \\
  0 \\
  p \\
\end{array}%
\right)  =  \left(%
\begin{array}{c}
  \sqrt{p^{2}+m^{2}}\cosh\chi \\
  \sqrt{p^{2}+m^{2}}\sinh\chi \\
  0 \\
  p \\
\end{array}%
\right), \nonumber \\
 R_{y}(\hat{p}_{\Lambda})
 & = &  \left(%
\begin{array}{ccc}
  \cos\theta_{\Lambda} & 0 & \sin\theta_{\Lambda} \\
  0 & 1 & 0 \\
  -\sin\theta_{\Lambda} & 0 & \cos\theta_{\Lambda} \\
\end{array}%
\right), \nonumber
\\
R_{y}(\hat{p}_{\Lambda P}) & = & R_{y}(\pi-\theta_{\Lambda}),
\nonumber
\end{eqnarray}
where $  E_{p}=\sqrt{p^{2}+m^{2}},$ and
 $ \tan\theta_{\Lambda}=(E_{p}\sinh\chi)/p=\frac{\sinh\chi}{\tanh\xi}.$
 \\
Thus,
%
%
\begin{eqnarray}
R_{y}(\hat{p}_{\Lambda})~\vec{\sigma} &= & \left(%
\begin{array}{ccc}
  \cos\theta_{\Lambda} & 0 & \sin\theta_{\Lambda} \\
  0 & 1 & 0 \\
  -\sin\theta_{\Lambda} & 0 & \cos\theta_{\Lambda} \\
\end{array}%
\right)\left(%
\begin{array}{c}
  \sigma_{x} \\
  \sigma_{y} \\
  \sigma_{z} \\
\end{array}%
\right) =\left(%
\begin{array}{c}
  \sigma_{x}\cos\theta_{\Lambda}+\sigma_{z}\sin\theta_{\Lambda} \\
  \sigma_{y} \\
  -\sigma_{x}\sin\theta_{\Lambda}+\sigma_{z}\cos\theta_{\Lambda} \\
\end{array}%
\right),
\end{eqnarray}
and
%
%
\begin{eqnarray}
a_{\Lambda p} & = & [R_{y}(\theta_{\Lambda})L_{z}(\eta)]^{-1}~a
    \nonumber\\
  & =  &  \left(%
\begin{array}{cccc}
  \cosh\eta & 0 & 0 &-\sinh\eta \\
  0 & 1 & 0 & 0 \\
  0 & 0 & 1 & 0 \\
  -\sinh\eta & 0 & 0 & \cosh\eta\\
\end{array}%
\right)\left(%
\begin{array}{cccc}
  1 & 0 & 0 & 0 \\
  0 & \cos\theta_{\Lambda} & 0 & -\sin\theta_{\Lambda} \\
  0 & 0 & 1 & 0\\
  0 & \sin\theta_{\Lambda} & 0 & \cos\theta_{\Lambda} \\
\end{array}%
\right)\left(%
\begin{array}{c}
  0 \\
  a_{x}\\
  a_{y}\\
  a_{z}\\
\end{array}%
\right), \label{ce23}
\end{eqnarray}
where $\tanh\eta=|\vec{p}_{\Lambda}|/E_{\Lambda
p}=\frac{\sqrt{\tanh^{2}\xi+\sinh^{2}}\chi}{\cosh\chi}$.
Thus the spatial part of $a_{\Lambda p}$ and its magnitude are
given by
%
%
\begin{eqnarray}
\vec{a}_{\Lambda p} & = &
(a_{x}\cos\theta_{\Lambda}-a_{z}\sin\theta_{\Lambda},a_{y},
  \cosh\eta(a_{x}\sin\theta_{\Lambda}+a_{z}\cos\theta_{\Lambda})), \nonumber \\
  |\vec{a}_{\Lambda p}| & = &  \sqrt{1+\sinh^{2}\eta ~(a_{x}\sin
\theta_{\Lambda}+a_{z}\cos\theta_{\Lambda})^{2}}.
\end{eqnarray}
Similarly from
 $b_{\Lambda
 p}=[R_{y}(\pi-\theta_{\Lambda})B_{z}(\eta)]^{-1}~b$,
we get
%
%
\begin{eqnarray}
\vec{b}_{\Lambda p} & = &
(-b_{x}\cos\theta_{\Lambda}-b_{z}\sin\theta_{\Lambda},b_{y},
\cosh\eta(b_{x}\sin\theta_{\Lambda}-b_{z}\cos\theta_{\Lambda})),
\nonumber \\
 |\vec{b}_{\Lambda Pp}| & = & \sqrt{1+\sinh^{2}\eta ~(-b_{x}\sin
\theta_{\Lambda}+b_{z}\cos\theta_{\Lambda})^{2}},
\end{eqnarray}
and
%
%
\begin{equation}
R_{y}(\hat{p}_{\Lambda P})~\vec{\sigma}=\left(%
\begin{array}{ccc}
  -\cos\theta_{\Lambda} & 0 & \sin\theta_{\Lambda} \\
  0 & 1 & 0 \\
  -\sin\theta_{\Lambda} & 0 & -\cos\theta_{\Lambda} \\
\end{array}%
\right)\left(%
\begin{array}{c}
  \sigma_{x} \\
  \sigma_{y} \\
  \sigma_{z} \\
\end{array}%
\right) =\left(%
\begin{array}{c}
  -\sigma_{x}\cos\theta_{\Lambda}+\sigma_{z}\sin\theta_{\Lambda} \\
  \sigma_{y} \\
  -\sigma_{x}\sin\theta_{\Lambda}-\sigma_{z}\cos\theta_{\Lambda} \\
\end{array}%
\right).
\end{equation}
Therefore, the tensor product of relativistic spin observables of
particle 1 and 2 for the joint spin measurement
 can be expressed as
%
%
\begin{equation}
\hat{a} \otimes \hat{b}=\frac{\vec{a}_{\Lambda p}\cdot
R(\vec{p}_{\Lambda}) ~\vec{\sigma}}{|\lambda(\vec{a}_{\Lambda
p}\cdot \vec{\sigma})|} \otimes \frac{\vec{b}_{\Lambda Pp}\cdot
R(\vec{p}_{\Lambda P}) ~\vec{\sigma}}{|\lambda(\vec{b}_{\Lambda
Pp}\cdot \vec{\sigma})|}
 \equiv   \frac{\vec{A}\cdot \vec{\sigma}
\otimes \vec{B}\cdot \vec{\sigma}}{|\vec{a}_{\Lambda
p}||\vec{b}_{\Lambda Pp}|} \label{ce27}
\end{equation}
where
%
%
\begin{eqnarray}
\vec{A} & = & \left(%
\begin{array}{c}
  a_{x}(\cos^{2}\theta_{\Lambda}-\cosh\eta\sin^{2}\theta_{\Lambda})-
  a_{z}(1+\cosh\eta)\sin\theta_{\Lambda}\cos\theta_{\Lambda}\\
  a_{y} \\
  a_{x}(1+\cosh\eta)\sin\theta_{\Lambda}\cos\theta_{\Lambda}-
  a_{z}(\sin^{2}\theta_{\Lambda}-\cosh\eta\cos^{2}\theta_{\Lambda}) \\
\end{array}%
\right), \nonumber\\
\vec{B} & = &
\left(%
\begin{array}{c}
  b_{x}(\cos^{2}\theta_{\Lambda}-\cosh\eta\sin^{2}\theta_{\Lambda})
  +b_{z}(1+\cosh\eta)\sin\theta_{\Lambda}\cos\theta_{\Lambda}\\
  b_{y} \\
  -b_{x}(1+\cosh\eta)\sin\theta_{\Lambda}\cos\theta_{\Lambda}
  -b_{z}(\sin^{2}\theta_{\Lambda}-\cosh\eta\cos^{2}\theta_{\Lambda}) \\
\end{array}%
\right).
\label{ce28}
\end{eqnarray}
Using the following relations
%
%
\begin{eqnarray}
\sigma_{x} \otimes \sigma_{x}~ |\Phi> & = & -\cos
\Omega_{p}~|\Psi_{\Lambda}^{-}> + \sin
\Omega_{p}~|\Phi_{\Lambda}^{+}>, \nonumber\\
\sigma_{y} \otimes \sigma_{y}~ |\Phi> & = & -\cos
\Omega_{p}~|\Psi_{\Lambda}^{-}> - \sin
\Omega_{p}~|\Phi_{\Lambda}^{+}>, \nonumber\\
\sigma_{z} \otimes \sigma_{z}~ |\Phi> &= & -\cos
\Omega_{p}~|\Psi_{\Lambda}^{-}> + \sin
\Omega_{p}~|\Phi_{\Lambda}^{+}>, \nonumber\\
\sigma_{x} \otimes \sigma_{z}~ |\Phi> & = & -\cos
\Omega_{p}~|\Phi_{\Lambda}^{+}> - \sin
\Omega_{p}~|\Psi_{\Lambda}^{-}>, \nonumber\\
\sigma_{z} \otimes \sigma_{x}~ |\Phi> & = & \cos
\Omega_{p}~|\Phi_{\Lambda}^{+}> + \sin
\Omega_{p}~|\Psi_{\Lambda}^{-}>,
\label{ce29}
\end{eqnarray}
and since the remaining terms do not contribute to the expectation
value, we finally get the following expression for the expectation
value of the joint spin measurement for the spin singlet.
%
%
\begin{eqnarray}
 & & <\hat{a} \otimes \hat{b}>      \nonumber \\
 & & =  \frac{-1}{|\vec{a}_{\Lambda
p}||\vec{b}_{\Lambda Pp}|}[(A_{x}B_{x}+A_{z}B_{z})\cos 2
~\Omega_{p}+A_{y}B_{y}+(A_{x}B_{z}-A_{z}B_{x})\sin2~\Omega_{p}]
\label{ce30}
\end{eqnarray}
Here, we examine two limiting cases of the above formula.
\\
1) When $\chi \rightarrow 0$, $<\hat{a} \otimes \hat{b}>
\rightarrow
\frac{-1}{\sqrt{1+a_{z}^{2}\sinh^{2}\xi}\sqrt{1+b_{z}^{2}\sinh^{2}\xi}}
(a_{x}b_{x}+a_{y}b_{y}+a_{z}b_{z}\cosh^{2}\xi)$.
\\
2) When $\xi \rightarrow 0$,
 $<\hat{a} \otimes \hat{b}>
\rightarrow
\frac{-1}{\sqrt{1+a_{x}^{2}\sinh^{2}\chi}\sqrt{1+b_{x}^{2}\sinh^{2}\chi}}
(a_{x}b_{x}\cosh^{2}\chi+a_{y}b_{y}+a_{z}b_{z})$.
\\
Notice that the second case exactly corresponds to the
Czachor's set-up and yields the same result.\\

Now, let us evaluate the Bell observable for a set of measurement
vectors which yield the maximal violation of the Bell's inequality
in the non-relativistic case:
%
%
\begin{eqnarray}
\vec{a}& = & (0,\frac{1}{\sqrt{2}}~,~\frac{1}{\sqrt{2}})~,~
\vec{a'}=(0,-\frac{1}{\sqrt{2}}~,~\frac{1}{\sqrt{2}}),
\nonumber\\
\vec{b} & = & (0~,~0~,~1) ~,~~~~ \vec{b'}=(0~,~1~,~0).
 \label{ce31}
\end{eqnarray}
With this set of measurement vectors, the Bell observable
$C(a,a',b,b')$ is given by
%
%
\begin{eqnarray}
C(a,a',b,b') &= & <\hat{a} \otimes \hat{b}>+<\hat{a} \otimes
\hat{b'}>+<\hat{a'} \otimes \hat{b}>-<\hat{a'} \otimes \hat{b'}>
~ \nonumber \\
&  =  & - \frac{2}{\sqrt{1+\sin^{2}\theta_{\Lambda}+\cosh^{2}
\eta\cos^{2}\theta_{\Lambda}}}  \nonumber \\
&  &  - \frac{2}{\sqrt{1+\sin^{2}\theta_{\Lambda}+\cosh^{2}
\eta\cos^{2}\theta_{\Lambda}}\sqrt{\sin^{2}\theta_{\Lambda}+\cosh^{2}
\eta\cos^{2}\theta_{\Lambda}}}  \nonumber \\
& &
 ~  \times
\{\left[(\cosh\eta\cos^{2}\theta_{\Lambda}-\sin^{2}\theta_{\Lambda})^{2}
-(1+\cosh\eta)^{2}\sin^{2}\theta_{\Lambda}
\cos^{2}\theta_{\Lambda}\right]\cos 2 \Omega_{p}
  \nonumber \\
&  &  ~~~~~
 -(1+\cosh\eta)(\cosh\eta\cos^{2}\theta_{\Lambda}
 -\sin^{2}\theta_{\Lambda})\sin
 2\theta_{\Lambda}\sin2\Omega_{p}\}.
 \label{ce32}
\end{eqnarray}
 Here also, we consider two limiting cases of the above formula.
\\
1) When $ \chi \rightarrow 0 , ~ \theta_{\Lambda} \rightarrow 0, ~
\eta \rightarrow \xi$, we get
 $|C(a,a',b,b')| \rightarrow
2(1+\cosh\xi)/\sqrt{2+\sinh^{2}\xi}$.
\\
2) When $\xi \rightarrow 0 , ~ \theta_{\Lambda} \rightarrow \pi/2,
~ \eta \rightarrow \chi$, we get $|C(a,a',b,b')| \rightarrow
2\sqrt{2}$.
\\
The first case is similar to the Czachor's set-up in the sense
that the observers are at rest and only the particles are moving
in opposite directions with the same speed. The result is the same
one that we can infer from the Czachor's result. The second case
corresponds to a case in the Czachor's in which the spin
measurement directions are perpendicular to the particles
movement. The result agrees with the Czachor's. \\

What happens if the two particles have different velocities not in
the opposite directions? In order to make the discussion simple we
consider the case when the observer is at rest in the lab frame.
Let  $p$ and $q$ be the momentum of particle 1 and 2,
respectively. In this case, the Wigner angle $\Omega_p$ is zero.
Then, the expectation value of the joint spin measurement is given
by
\begin{equation}
<\hat{a} \otimes \hat{b}>=<\Psi|\frac{\vec{a}\cdot
R(\vec{p})\vec{\sigma}}{|\lambda(\vec{a}_{p}\cdot \vec{\sigma})|}
\otimes \frac{\vec{b}\cdot
R(\vec{q})\vec{\sigma}}{|\lambda(\vec{b}_{q}\cdot \vec{\sigma})|}
|\Psi>=<\Psi|\frac{\vec{A}\cdot \vec{\sigma}}{|\vec{a}_{p}|}
\otimes \frac{\vec{B}\cdot \vec{\sigma}}{|\vec{b}_{q}|} |\Psi>
\end{equation}
where
\begin{eqnarray}
 \vec{A} &= & \left(%
\begin{array}{c}
  a_{x}(\cos^{2}\theta_{p}-\cosh\xi_{p}\sin^{2}\theta_{p})-a_{z}(1+\cosh\xi_{p})\sin\theta_{p}\cos\theta_{p} \\
  a_{y}\\
   a_{x}(1+\cosh\xi_{p})\sin\theta_{p}\cos\theta_{p}-a_{z}(\sin^{2}\theta_{p}-\cosh\xi_{p}\cos^{2}\theta_{p})\\
\end{array}%
\right) , \nonumber \\
\vec{B} & = & \left(%
\begin{array}{c}
  b_{x}(\cos^{2}\theta_{q}-\cosh\xi_{q}\sin^{2}\theta_{q})-b_{z}(1+\cosh\xi_{q})\sin\theta_{q}\cos\theta_{q} \\
  b_{y}\\
   b_{x}(1+\cosh\xi_{q})\sin\theta_{q}\cos\theta_{q}-b_{z}(\sin^{2}\theta_{q}-\cosh\xi_{q}\cos^{2}\theta_{q})\\
\end{array}
\right) , \label{cea1}
\end{eqnarray}
and its value becomes
\begin{equation}
<\hat{a}\otimes \hat{b}>=-\frac{\vec{A}\cdot
\vec{B}}{|\vec{a}_{p}||\vec{b}_{q}|} ~ .
\end{equation}
For the same set of measurement vectors as in (\ref{ce31}), the
Bell observable is given by
\begin{eqnarray}
C(a,a',b,b')& = & -\frac{2}{\sqrt{2+\sinh^{2}\xi_{p}\cos^{2}\theta_{p}}}\nonumber\\
& & -\frac{2}
{\sqrt{2+\sinh^{2}\xi_{p}\cos^{2}\theta_{p}}\sqrt{1+\sinh^{2}\xi_{q}\cos^{2}\theta_{q}}}
 \nonumber \\
 & & ~ \times
  \left\{
  (1+\cosh\xi_{p})(1+\cosh\xi_{q})\sin_{\theta_{p}}\cos_{\theta_{p}}\sin_{\theta_{q}}\cos_{\theta_{p}}
  \right. \nonumber \\
  & & ~~~~~ \left.
+(\sin^{2}{\theta_{p}}-\cosh\xi_{p}\cos^{2}\theta_{p})
(\sin^{2}{\theta_{q}}-\cosh\xi_{q}\cos^{2}\theta_{q}) \right\} .
\end{eqnarray}
One can check that this result reduces to the one from
(\ref{ce32}),  if $\vec{p}$
 and $\vec{q}$ are in the opposite directions with the same
 magnitude.
In the next section, we shall see that we can find the corrected
vector set for the maximal violation of Bell's inequality in this
case also.
\\
%
%
%
%
%
%
%
%
%

\section*{IV. Corrected Bell observable for spin singlet }

In this section, we will show that by appropriately choosing the
vector set for spin measurements the maximal violation of the
Bell's inequality can be achieved even in a relativistically
moving inertial frame. For the non-relativistic case, it is known
that a fully entangled state such as the spin singlet maximally
violates the Bell's inequality giving the value of the Bell
observable $2 \sqrt{2} $. For the spin singlet case, the vector
set inducing the maximal violation may be chosen as
%
%
\begin{eqnarray}
\vec{a} & = & (0,\frac{1}{\sqrt{2}}~,~\frac{1}{\sqrt{2}})~,~
\vec{a'}=(0,-\frac{1}{\sqrt{2}}~,~\frac{1}{\sqrt{2}})
\nonumber\\
\vec{b}& = & (0~,~0~,~1) ~,~~~ \vec{b'}=(0~,~1~,~0).
\label{ce33}
\end{eqnarray}
In the non-relativistic case, the expectation value of the joint
spin measurement for the spin singlet is given by
%
%
\begin{equation}
<\hat{a} \otimes \hat{b}> = -\vec{a} \cdot \vec{b} \label{ev-sing}
\end{equation}
for a set of measurement vectors, $\vec{a}$ for particle 1 and
$\vec{b}$ for particle 2, where $\hat{a}=\vec{a}\cdot
\vec{\sigma},~ \hat{b}=\vec{b}\cdot \vec{\sigma}$.
However, when the movement of the particles or the observers
become relativistic the above expectation value is not maintained
as Czachor and others have shown \cite{czachor,tu,ahn}.

 Following the reasoning of Terashima and Ueda \cite{tu}, here we
investigate whether we can find a set of spin measurement
directions which preserve the non-relativistic expectation value
 (\ref{ev-sing}) even under relativistic situations.
In order to do this, we consider the case  in which a new set of
spin measurement directions $\vec{a}_{c}, ~ \vec{b}_{c}$ in a
Lorentz boosted frame yields the relation (\ref{ev-sing})
%
%
\begin{equation}
<\hat{a}_{c} \otimes \hat{b}_{c}> = -\vec{a} \cdot \vec{b}
\label{corr-set}
\end{equation}
with the previously chosen vector set $ \vec{a}, \vec{b} $ in the
non-relativistic lab frame.
The existence of the new vector set $\vec{a}_{c}, ~ \vec{b}_{c}$
implies that in the new frame the correlation between the two
entangled particles can be seen when the measurement is performed
along these new direction vectors not along the previously given
directions $ \vec{a}, \vec{b} $ in the lab frame.
Thus we will try to find $\vec{a}_{c}, ~ \vec{b}_{c}$ satisfying
(\ref{corr-set}) for a simple case of Czachor, then for a more
general case.

In the Czachor's set-up in which the both particles are moving in
the $+z$ direction, the relation (\ref{corr-set}) is satisfied if
\[ \frac{\vec{a}_{c}\cdot
\vec{\sigma}}{|\vec{a}_{c}|}=\vec{a}\cdot \vec{\sigma}. \]
Let us denote $\vec{a}_{c}=(a_{cx},a_{cy},a_{cz})$, then from the
relation $ ~ L_{z}(-\xi)a_{c}=a_{cp} ~, ~ \tanh \xi = \beta_p ~$,
%
%
\begin{equation}
\left(%
\begin{array}{cccc}
  \cosh\xi & 0 & 0 & -\sinh\xi \\
  0 & 1 & 0 & 0 \\
  0 & 0 & 1 & 0 \\
  -\sinh\xi & 0 & 0 & \cosh\xi \\
\end{array}%
\right)\left(%
\begin{array}{c}
  0 \\
  a_{cx} \\
  a_{cy} \\
  a_{cz} \\
\end{array}%
\right)=\left(%
\begin{array}{c}
  -a_{cz}\sinh\xi \\
  a_{cx} \\
  a_{cy} \\
  a_{cz}\cosh\xi \\
\end{array}%
\right), \label{ce36}
\end{equation}
thus we can write $\vec{a}_{cp}=(a_{cx},a_{cy},a_{cz}\cosh\xi)$,
and we get the following equation for the corrected vector
$\vec{a}_{c}$.
%
%
\begin{equation}
\frac{1}{\sqrt{1+a_{cz}^{2}\sinh^{2}\xi}}(a_{cx},a_{cy},a_{cz}\cosh\xi)=(a_{x},a_{y},a_{z})
\end{equation}
Solving the above equation, we get
%
%
\begin{equation}
a_{cx}=\frac{a_{x}}{\sqrt{1-a_{z}^{2}\tanh^{2}\xi}} , ~~
a_{cy}=\frac{a_{y}}{\sqrt{1-a_{z}^{2}\tanh^{2}\xi}}, ~~
a_{cz}=\frac{a_{z}}{\cosh\xi\sqrt{1-a_{z}^{2}\tanh^{2}\xi}}.
\label{ce38}
\end{equation}
Similarly, we get the same expression for the remaining corrected
 vector $\vec{b}_{c}$ of the new frame.
 The expectation value
of the joint spin measurement for the corrected vector set is now
given by (\ref{ce14}),
%
%
\begin{eqnarray}
<\hat{a_{c}} \otimes \hat{b_{c}}> &
 = & -\frac{(a_{cx}b_{cx}+a_{cy}b_{cy}+a_{cz}b_{cz}\cosh^{2}\xi)}
{\sqrt{1+a_{cz}^{2}\sinh^{2}\xi}\sqrt{1+b_{cz}^{2}\sinh^{2}\xi}}
\label{ce39} \\
& = & -(a_{x}b_{x}+a_{y}b_{y}+a_{z}b_{z}) = -\vec{a}\cdot\vec{b},
\nonumber
\end{eqnarray}
and thus satisfies our requirement.

For the evaluation of the Bell observable, we first evaluate the
corrected vectors for $ \vec{a}, \vec{a}'$ given by (\ref{ce33})
by use of (\ref{ce38})and similarly for $ \vec{b}, \vec{b}'$ given
by (\ref{ce33}):
%
%
\begin{eqnarray}
\vec{a}_{c} &= &(0,\frac{1}{\sqrt{2-\tanh^{2}\xi}}~,
~\frac{1}{\cosh\xi\sqrt{2-\tanh^{2}\xi}})~,~\nonumber\\
\vec{a'}_{c} & = &
(0,\frac{-1}{\sqrt{2-\tanh^{2}\xi}}~,~\frac{1}{\cosh\xi\sqrt{2-\tanh^{2}\xi}})~,~
\nonumber\\
\vec{b}_{c}& = & (0~,~0~,~1) ~, ~~~~~ \vec{b'}_{c}=(0~,~1~,~0).
\end{eqnarray}
By use of the formula (\ref{ce39}) the Bell observable with the
above corrected vector set is now evaluated as
%
%
\begin{eqnarray}
C(a_{c},a'_{c},b_{c},b'_{c}) & = & <\hat{a}_{c}\otimes
\hat{b}_{c}>+<\hat{a}_{c}\otimes
\hat{b'}_{c}>+<\hat{a'}_{c}\otimes
\hat{b}_{c}>-<\hat{a'}_{c}\otimes \hat{b'}_{c}> \nonumber\\
& = &
-\frac{1}{\sqrt{2}}-\frac{1}{\sqrt{2}}-\frac{1}{\sqrt{2}}-\frac{1}{\sqrt{2}}
=-2\sqrt{2}
\end{eqnarray}
retrieving the value of the maximal violation of the Bell's
inequality in the non-relativistic case.
\\

%
%
%

%
Next, we consider a more general case as in section III. The two
particles of the spin singlet are moving in the $+z$ and $-z$
directions respectively in the lab frame, and the observers, Alice
and Bob, are sitting in a moving frame which is Lorentz boosted
toward the $-x$ direction.
In this case, the expectation value of the joint spin measurement
for a corrected vector set is given by (\ref{ce30}),
%
%
\begin{eqnarray}
 & & <\hat{a}_{c} \otimes \hat{b}_{c}> \label{ce42}  \\
 &  &  =\frac{-1}{|\vec{a}_{c\Lambda
p}||\vec{b}_{c\Lambda Pp}|}[(A_{cx}B_{cx}+A_{cz}B_{cz})\cos 2
~\Omega_{p}+A_{cy}B_{cy}+(A_{cx}B_{cz}-A_{cz}B_{cx})\sin2~\Omega_{p}],
\nonumber
\end{eqnarray}
where $\vec{A}_{c} ,\vec{B}_{c}$ correspond to $\vec{A} ,\vec{B}$
of (\ref{ce28}) in which $\vec{a} ,\vec{b}$ are replaced with
$\vec{a}_{c} ,\vec{b}_{c}$, and other notations such as $\Omega_p,
\xi, \chi, \eta,  \theta_\Lambda , $ etc. are the same as in
section III.

Here again, we will get the corrected vector set of spin
measurement directions if we find $\vec{a}_{c} ,\vec{b}_{c}$ which
give the expectation value (\ref{ce42}) to be $- \vec{a} \cdot
\vec{b} $. Namely, we want to find $\vec{a}_{c} ,\vec{b}_{c}$ that
satisfy the following equation:
%
%
\begin{equation}
 \frac{-1}{|\vec{a}_{c\Lambda
p}||\vec{b}_{c\Lambda Pp}|}[(A_{cx}B_{cx}+A_{cz}B_{cz})\cos 2
~\Omega_{p}+A_{cy}B_{cy}+(A_{cx}B_{cz}-A_{cz}B_{cx})\sin2~\Omega_{p}]
= - \vec{a} \cdot \vec{b}
\label{ce43}
\end{equation}
where $(\vec{a}_{c\Lambda p} ,\vec{b}_{c\Lambda Pp})$ correspond
to $(\vec{a}_{\Lambda p} ,\vec{b}_{\Lambda Pp})$ of the previous
section as above.
One can check that the equation (\ref{ce43}) is satisfied if the
following relation is satisfied
%
%
\begin{equation}
\frac{A_{ci}}{|\vec{a}_{c\Lambda p}|}=\bar{a}_i ~, ~~
\frac{B_{ci}}{|\vec{b}_{c\Lambda Pp}|}=\bar{b}_i ~~~ \text{for} ~~
i=(x,y,z),
 \label{ce44}
\end{equation}
where
%
%
\begin{equation}
\bar{a}_i \equiv  R_y (\Omega_p) a_i ~, ~~ \bar{b}_i \equiv  R_y
(- \Omega_p) b_i  ~~~ \text{with}~~
R_{y}(\Omega_{p})=\left(%
\begin{array}{ccc}
  \cos \Omega_{p} & 0 & \sin\Omega_{p} \\
  0 & 1 & 0 \\
  -\sin\Omega_{p} & 0 & \cos\Omega_{p} \\
\end{array}%
\right). \label{ce45}
\end{equation}
Solving (\ref{ce44}), we obtain $\vec{a}_{c} ,\vec{b}_{c}$ in
terms of $\vec{a} ,\vec{b}$:
%
%
\begin{eqnarray}
a_{cz}&=&\frac{\overline{a}_{z}}
{\sqrt{\left[F_{a}(1+\cosh\eta)\sin\theta_{\Lambda}\cos\theta_{\Lambda}-
(\sin^{2}\theta_{\Lambda}-\cosh\eta\cos^{2}\theta_{\Lambda})\right]^{2}
 -\overline{a}_{z}^{2}\sinh^{2}\eta\left(
F_{a}\sin\theta_{\Lambda}+\cos\theta_{\Lambda}\right)^{2}}},
\nonumber\\
a_{cx}&=&\overline{a}_{x}\sqrt{1+a_{cz}^{2}\sinh^{2}\eta\left(F_{a}\sin\theta_{\Lambda}
+\cos\theta_{\Lambda}\right)^{2}},
\nonumber\\
a_{cy}&=&a_{y}\sqrt{1+a_{cz}^{2}\sinh^{2}\eta\left(F_{a}\sin\theta_{\Lambda}
+\cos\theta_{\Lambda}\right)^{2}}, \label{ce46}
\\
b_{cz}
 &=&\frac{\bar{b}_{z}}
 {\sqrt{\left[F_{b}(1+\cosh\eta)\sin\theta_{\Lambda}\cos\theta_{\Lambda}-
 (\sin^{2}\theta_{\Lambda}-\cosh\eta\cos^{2}\theta_{\Lambda})\right]^{2}
 -\bar{b}_{z}^{2}\sinh^{2}\eta\left(F_{b}\sin\theta_{\Lambda}
 -\cos\theta_{\Lambda}\right)^{2}}},
\nonumber\\
b_{cx}&=&\overline{b}_{x}\sqrt{1+b_{cz}^{2}\sinh^{2}\eta\left(F_{b}\sin\theta_{\Lambda}
 -\cos\theta_{\Lambda}\right)^{2}},
 \nonumber\\
b_{cy}&= &
   b_{y}\sqrt{1+b_{cz}^{2}\sinh^{2}\eta\left(F_{b}\sin\theta_{\Lambda}
   -\cos\theta_{\Lambda}\right)^{2}},
\nonumber
\end{eqnarray}
where $ \bar{a}_i ~, ~\bar{b}_i ~~\text{for}~~ i=x,y,z ~$ are
given by (\ref{ce45}), and
\begin{eqnarray}
F_{a} & =
   & \frac{(1+\cosh\eta)\tan\theta_{\Lambda}-f_{a}(\tan^{2}\theta_{\Lambda}-\cosh\eta)
}{(1-\cosh\eta\tan^{2}\theta_{\Lambda})-f_{a}(1+\cosh\eta)\tan\theta_{\Lambda}},
\nonumber \\
F_{b} & =
  & -\frac{(1+\cosh\eta)\tan\theta_{\Lambda}+f_{b}(\tan^{2}\theta_{\Lambda}-\cosh\eta)
}{(1-\cosh\eta\tan^{2}\theta_{\Lambda})+f_{b}(1+\cosh\eta)\tan\theta_{\Lambda}},
\nonumber \\
\text{with} &   &  f_{a}  \equiv
\frac{\overline{a}_{x}}{\overline{a}_{z}}~,~~~
 f_{b} \equiv \frac{\overline{b}_{x}}{\overline{b}_{z}}. \nonumber
\end{eqnarray}
And thus $ |\vec{a}_{c\Lambda p}|, ~ |\vec{b}_{c\Lambda Pp}| $ are
given by
\begin{eqnarray}
|\vec{a}_{c\Lambda p}|&
=&\sqrt{1+a_{cz}^{2}\sinh^{2}\eta\left(F_{a}\sin\theta_{\Lambda}
+\cos\theta_{\Lambda}\right)^{2}},
\nonumber\\
|\vec{b}_{c\Lambda Pp}|
&=&\sqrt{1+b_{cz}^{2}\sinh^{2}\eta\left(F_{b}\sin\theta_{\Lambda}
-\cos\theta_{\Lambda}\right)^{2}} . \nonumber
\end{eqnarray}
Now, we consider how the correlation due to entanglement is
changed by Lorentz boost. For the spin singlet the two spins of
particle 1 and 2 are always antiparallel in the non-relativistic
case. Thus we would like to see how the corrected vector set in
the Lorentz boosted moving frame shows the correlation between the
two spins of the entangled particles that exists in the
non-relativistic lab frame.

As expressed in (\ref{ce43}), the corrected vector set for the
spin singlet is defined to satisfy
\[
<\hat{a}_{c} \otimes \hat{b}_{c}> =-\vec{a} \cdot \vec{b}.
\]
In other words, when the two measurement directions for particle 1
and 2 are the same, $\vec{a} = \vec{b}$, in the non-relativistic
lab frame, the expectation value of the joint spin measurement
with the new directions, $( \vec{a}_{c} ~, ~  \vec{b}_{c})$, in
the Lorentz boosted moving frame should be $-1$. Here, we would
like to see what these corrected spin measurement directions are
and consider the meaning of these new directions when $\vec{a} =
\vec{b}=(0,0,1)$.
In this case, from (\ref{ce45}) $~  \bar{a}_i ~, ~ \bar{b}_i  ~ ~
\text{for} ~~ i=x,y,z $ are given by
\[
\{ \bar{a}_{i} \} = (  \sin\Omega_{p}, 0 , \cos\Omega_{p} ) ~,~~
 \{ \bar{b}_{i} \} = (  -\sin\Omega_{p},  0 , \cos\Omega_{p} ) ,
\]
and thus from (\ref{ce46}) the corrected vectors are given as
follows.
\begin{eqnarray}
a_{cz}&=&\frac{\cos \Omega_{p}} {\sqrt{D_a}},
\nonumber\\
a_{cx}&=&\sin\Omega_{p}\sqrt{1+\cos^{2}\Omega_{p}\sinh^{2}\eta\left(F_{a}\sin\theta_{\Lambda}
+\cos\theta_{\Lambda}\right)^{2}},
\nonumber\\
a_{cy}&=&0 , \nonumber \\
 b_{cz}
 &=&\frac{\cos\Omega_{p}} {\sqrt{D_b}} ,
 \nonumber\\
b_{cx}&=&-\sin\Omega_{p}\sqrt{1+\cos^{2}\Omega_{p}\sinh^{2}\eta\left(F_{a}\sin\theta_{\Lambda}
+\cos\theta_{\Lambda}\right)^{2}} ,
\nonumber\\
b_{cy}&=&0, \nonumber
\end{eqnarray}
where
\[
D_a =
\left[F_{a}(1+\cosh\eta)\sin\theta_{\Lambda}\cos\theta_{\Lambda}-
(\sin^{2}\theta_{\Lambda}-\cosh\eta\cos^{2}\theta_{\Lambda})\right]^{2}
 -\cos^{2}\Omega_{p}\sinh^{2}\eta\left(
F_{a}\sin\theta_{\Lambda}+\cos\theta_{\Lambda}\right)^{2} ,
\]
\[
D_b =
 \left[F_{b}(1+\cosh\eta)\sin\theta_{\Lambda}\cos\theta_{\Lambda}+
(\sin^{2}\theta_{\Lambda}-\cosh\eta\cos^{2}\theta_{\Lambda})\right]^{2}
-\cos^{2}\Omega_{p} \sinh^{2}\eta\left(
 F_{a}\sin\theta_{\Lambda}+\cos\theta_{\Lambda}\right)^{2} ,
\]
and
\[
F_{a} = \frac{(1+\cosh\eta)\tan\theta_{\Lambda} -\tan
\Omega_{p}(\tan^{2}\theta_{\Lambda}-\cosh\eta)
}{(1-\cosh\eta\tan^{2}\theta_{\Lambda})-\tan
\Omega_{p}(1+\cosh\eta)\tan\theta_{\Lambda}} ,
\]
with
$ ~  \tanh \xi = \beta_p , \ \tanh  \chi = \beta_{\Lambda} , \
\cosh\eta=\cosh\xi\cosh\chi , \ \tan\theta_{\Lambda}
=\frac{\sinh\chi}{\tanh\xi} , \ \tan \Omega_{p}
=\frac{\sinh\xi\sinh\chi}{\cosh\xi+\cosh\chi} ~ .$
\\

In the limit, when
$\xi \rightarrow \infty ~,~ \chi\rightarrow \infty  $, the above
result yields after some numerical calculation
 \[F_{a}
\rightarrow 0~,~a_{cz} \rightarrow 0 , ~a_{cx} \rightarrow 1 , \]
and
\[F_{b} \rightarrow
0~,~b_{cz} \rightarrow 0 , ~b_{cx} \rightarrow -1 . \]
This result tells us that in the highly relativistic limit when
the boost speed reaches the speed of light both spins become
parallel not anti-parallel since the two spin measurement
directions should be opposite in order to maintain the same
expectation value $-1$ for the joint spin measurement in the
moving frame. This agrees with what we expected for spin rotation
under Lorentz boost.
\\

Finally, we consider the case that we dealt with at the end of the
last section in which the two particles are not moving in the
opposite directions. Namely, the two particles have arbitrary
momenta $p$ and $q$ and the observer is at rest in the lab frame.
Following the same argument, the corrected vector set should
satisfy the following condition.
\begin{equation}
<\hat{a_{c}} \otimes
\hat{b_{c}}>=-\frac{\vec{A}_{c}\cdot\vec{B}_{c}}{|\vec{a}_{c
p}||\vec{b}_{c q}|}=-\vec{a}\cdot \vec{b} ~ .
\end{equation}
Here, $\vec{A}_{c}, ~ \vec{B}_{c}$ are given by $\vec{A}, ~
\vec{B}$ in (\ref{cea1}) with $\vec{a}, ~ \vec{b}$ replaced with
$\vec{a}_{c}, ~ \vec{b}_{c}$, respectively.
This condition can be split into two conditions
\begin{equation}
\frac{\vec{A}_{c}}{|\vec{a}_{c p}|}=\vec{a} ,
 ~~ \frac{\vec{B}_{c}}{|\vec{b_{cq}|}}=\vec{b} ,
\end{equation}
and the result is given by for $\vec{a}_{c}~$,
\begin{eqnarray}
a_{cz}=\frac{a_{z}}{\sqrt{D_{a}}}
\end{eqnarray}
where
\[D_{a}=[F_{a}(1+\cosh\xi_{p})\sin\theta_{p}\cos\theta_{p}-(\sin^{2}\theta_{p}-\cosh\xi_{p}\cos^{2}\theta_{p})]^{2}
-a^{2}_{z}\sinh^{2}\xi_{p}(F_{a}\sin\theta_{p}+\cos\theta_{p})^{2},
\]
\[F_{a}=\frac{(1+\cosh\xi_{p})\tan\theta_{p}-\frac{a_{x}}{a_{z}}(\tan^{2}\theta_{p}-\cosh\xi_{p})}{(1-\cosh\xi_{p}\tan^{2}\theta_{p})-
\frac{a_{x}}{a_{z}}(1+\cosh\xi_{p})\tan\theta_{p}},
\]
and
\[a_{cx}=a_{x}\sqrt{1+a^{2}_{cz}\sinh^{2}\xi_{p}(F_{a}\sin\theta_{p}+\cos\theta_{p})^{2}},
\]
\[a_{cy}=a_{y}\sqrt{1+a^{2}_{cz}\sinh^{2}\xi_{p}(F_{a}\sin\theta_{p}+\cos\theta_{p})^{2}}
.
\]
In the case of $\vec{b}_{c}$, $
b_{cz}=\frac{b_{z}}{\sqrt{D_{b}}}$,
 and $\vec{a}$ and $\vec{p}$ are replaced with $\vec{b}$ and
$\vec{q}$, respectively, in the expression for $\vec{a}_{c}$.
One can check that this result reduces to the one from
(\ref{ce46}) with $\Omega_{p}=0, ~ \eta \rightarrow \xi, ~ \tan
\theta_{\Lambda}=0$, if $\vec{p}$
 and $\vec{q}$ are in the opposite directions with the same
 magnitude.
\\

%
%
%
%
%
%
%
%
%
%
%
%

\section*{V. Corrected Bell observable for the Bell states}

In this section, we will find corrected vector sets of spin
measurement directions for the remaining Bell states.

The Bell states are defined by \cite{nc}
%
%
\begin{eqnarray}
|\Phi_{p}^{(+)}> & \equiv & \frac{1}{\sqrt{2}}(|p,1/2>|-p,1/2>+|p,-1/2>|-p,-1/2>), \nonumber\\
|\Phi_{p}^{(-)}> & \equiv & \frac{1}{\sqrt{2}}(|p,1/2>|-p,1/2>-|p,-1/2>|-p,-1/2>), \nonumber\\
|\Psi_{p}^{(+)}> & \equiv &
\frac{1}{\sqrt{2}}(|p,1/2>|-p,-1/2>+|p,-1/2>|-p,1/2>),
 \label{ce47} \\
|\Psi_{p}^{(-)}> & \equiv &
\frac{1}{\sqrt{2}}(|p,1/2>|-p,-1/2>-|p,-1/2>|-p,1/2>), \nonumber
\end{eqnarray}
and transform under Lorentz boost as
%
%
\begin{eqnarray}
U(\Lambda)|\Phi_{p}^{(+)}>&=&\cos\Omega_{p} ~|\Phi_{\Lambda p}^{(+)}>
 - \sin\Omega_{p}~ |\Psi_{\Lambda p}^{(-)}>, \nonumber\\
U(\Lambda)|\Phi_{p}^{(-)}>&=&|\Phi_{\Lambda p}^{(-)}>, \nonumber\\
U(\Lambda)|\Psi_{p}^{(+)}>&=&|\Psi_{\Lambda p}^{(+)}>, \label{ce48} \\
U(\Lambda)|\Psi_{p}^{(-)}>&=&\cos\Omega_{p}~|\Psi_{\Lambda
p}^{(-)}>+\sin\Omega_{p}~|\Phi_{\Lambda p}^{(+)}>,    \nonumber
\end{eqnarray}
where  $\Omega_{p}$ is the Wigner angle due to the Lorentz boost
$\Lambda $ performed to a particle with momentum $\vec{p}$ and is
given by (\ref{ce21}).
In the non-relativistic case, the expectation values of the joint
spin measurement for the Bell states are given by
%
%
\begin{eqnarray}
<\Phi_{p}^{(+)}| ~\vec{a}\cdot\vec{\sigma} \otimes
\vec{b}\cdot\vec{\sigma}~| \Phi_{p}^{(+)}>
 & = & a_{x}b_{x}-a_{y}b_{y}+a_{z}b_{z}, \nonumber\\
<\Phi_{p}^{(-)}| ~\vec{a}\cdot\vec{\sigma} \otimes
\vec{b}\cdot\vec{\sigma}~| \Phi_{p}^{(-)}>
 & = & a_{x}b_{x}-a_{y}b_{y}+a_{z}b_{z}, \nonumber\\
<\Psi_{p}^{(+)}| ~\vec{a}\cdot\vec{\sigma} \otimes
\vec{b}\cdot\vec{\sigma}~| \Psi_{p}^{(+)}>
 & = & a_{x}b_{x}+a_{y}b_{y}-a_{z}b_{z}, \label{ce49} \\
<\Psi_{p}^{(-)}| ~\vec{a}\cdot\vec{\sigma} \otimes
\vec{b}\cdot\vec{\sigma}~| \Psi_{p}^{(-)}>
 & = & -a_{x}b_{x}-a_{y}b_{y}-a_{z}b_{z} ~. \nonumber
\end{eqnarray}

In the relativistic case, we consider the general case that we
studied in the previous sections in which particle 1 and 2 are
moving in the $+z$ and $-z$ directions respectively in the lab
frame, and the two observers for particle 1 and 2, Alice and Bob,
are Lorentz boosted to the $-x$ direction, the expectation values
of the joint spin measurement are given by use of the formula
(\ref{ce27}):
%
%
\begin{eqnarray}
<\tilde{\Phi}^{(+)}| \hat{a} \otimes \hat{b}| \tilde{\Phi}^{(+)}>
&=&\frac{1}{|\vec{a}_{\Lambda p}||\vec{b}_{\Lambda
Pp}|}[(A_{x}B_{x}+A_{z}B_{z})\cos 2 \Omega_{p} -A_{y}B_{y}
+(A_{x}B_{z}-A_{z}B_{x})\sin 2 \Omega_{p}],
\nonumber\\
<\tilde{\Phi}^{(-)}|\hat{a} \otimes \hat{b} | \tilde{\Phi}^{(-)}>
&=&\frac{1}{|\vec{a}_{\Lambda p}||\vec{b}_{\Lambda
Pp}|}[A_{x}B_{x}-A_{y}B_{y}+A_{z}B_{z}], \nonumber\\
<\tilde{\Psi}^{(+)}| \hat{a} \otimes \hat{b}| \tilde{\Psi}^{(+)}>
&=&\frac{1}{|\vec{a}_{\Lambda p}||\vec{b}_{\Lambda
Pp}|}[A_{x}B_{x}+A_{y}B_{y}-A_{z}B_{z}], \label{ce50}  \\
<\tilde{\Psi}^{(-)}| \hat{a} \otimes \hat{b}| \tilde{\Psi}^{(-)}>
&=&\frac{-1}{|\vec{a}_{\Lambda p}||\vec{b}_{\Lambda
Pp}|}\left[(A_{x}B_{x}+A_{z}B_{z})\cos 2 \Omega_{p} +A_{y}B_{y}
+(A_{x}B_{z}-A_{z}B_{x})\sin 2 \Omega_{p}\right], \nonumber
\end{eqnarray}
where $A_i ~, ~ B_i ~~\text{for} ~~ i=x,y,z~$ are given by
(\ref{ce28}), and we denoted the transformed Bell states as
\begin{eqnarray}
|\tilde{\Phi}^{(+)}> & = & U(\Lambda)|\Phi_{p}^{(+)}> ~, ~~
|\tilde{\Phi}^{(-)}>=U(\Lambda)|\Phi_{p}^{(-)}> , \nonumber \\
|\tilde{\Psi}^{(+)}> & = & U(\Lambda)|\Psi_{p}^{(+)}> ~, ~~
|\tilde{\Psi}^{(-)}> =U(\Lambda)|\Psi_{p}^{(-)}> .\nonumber
\end{eqnarray}
Here, we directly consider the corrected vector sets for maximal
violation of the Bell's inequality as we have done in the singlet
case. First we notice that for the states $ \{ |
\tilde{\Phi}^{(+)}> ~, ~ | \tilde{\Psi}^{(-)}>\}$ the corrected
vector set $(\vec{a}_{c}~, ~ \vec{b}_{c})$ should satisfy
%
%
\begin{eqnarray}
\frac{\vec{A}_{c}}{|\vec{a}_{c\Lambda p}|}&=& R_{y}(\Omega_{p})\vec{a}, \nonumber \\
\frac{\vec{B}_{c}}{|\vec{b}_{c\Lambda
Pp}|}&=&R_{y}(-\Omega_{p})\vec{b}, \label{ce51}
\end{eqnarray}
in order to give the same expectation values of the
non-relativistic case such that
%
%
\begin{eqnarray}
<\hat{a}_{c} \otimes \hat{b}_{c}>
&=&a_{x}b_{x}-a_{y}b_{y}+a_{z}b_{z}  ~~\text{for} ~~ | \tilde{\Phi}^{(+)}> , \nonumber \\
<\hat{a}_{c} \otimes \hat{b}_{c}>
&=&-a_{x}b_{x}-a_{y}b_{y}-a_{z}b_{z} ~~\text{for} ~~|
\tilde{\Psi}^{(-)}>. \label{ce52}
\end{eqnarray}
Next, for the states $ \{ | \tilde{\Phi}^{(-)}> , |
\tilde{\Psi}^{(+)}> \}$  the corrected vector set
$(\vec{a}_{c},\vec{b}_{c})$ should satisfy
%
%
\begin{eqnarray}
\frac{\vec{A}_{c}}{|\vec{a}_{c\Lambda p}|}&=& \vec{a},  \nonumber \\
\frac{\vec{B}_{c}}{|\vec{b}_{c\Lambda Pp}|}&=&\vec{b},
\label{ce53}
\end{eqnarray}
such that
%
%
\begin{eqnarray}
<\hat{a}_{c} \otimes \hat{b}_{c}>
&=&a_{x}b_{x}-a_{y}b_{y}+a_{z}b_{z} ~ , ~~ \text{for}~~ | \tilde{\Phi}^{(-)}> , \nonumber \\
<\hat{a}_{c} \otimes \hat{b}_{c}>
&=&a_{x}b_{x}+a_{y}b_{y}-a_{z}b_{z} ~, ~~ \text{for}~~
|\tilde{\Psi}^{(+)}> .   \label{ce54}
\end{eqnarray}
Here, $(\vec{A}_{c},~ \vec{B}_{c})$ are the one with
$(\vec{a}_{c},~ \vec{b}_{c})$ instead of $(\vec{a}, ~ \vec{b})$ in
(\ref{ce28}).

In this manner, we can find the corrected vector sets for the
other Bell states once the original vector sets which induce the
maximal violation of the Bell's inequality for each Bell state in
the non-relativistic case are given.
Namely, once we find the solutions for the equations (\ref{ce51})
and (\ref{ce53}), then we have the corrected vector sets for joint
spin measurement which will induce the maximal violation of the
Bell's inequality for the remaining Bell states.
\\

%
%
%
%
%
%
%
%
%
%
%
%
%

\section*{VI. Conclusion}

In this paper, we show that by appropriate rotations of the
directions of spin measurement  one can achieve the maximal
violation of the Bell's inequality even in a relativistically
moving frame if the state is fully entangled in a non-relativistic
lab frame.
In order to do this, we first define the relativistic spin
observable which we use for the spin measurement in an arbitrary
Lorentz boosted inertial frame. With this relativistic spin
observable we evaluate the expectation values of the joint spin
measurement for the Bell states in a Lorentz boosted frame.
In the spin singlet case, the expectation value evaluated in a lab
frame in which the two particles are moving in the same direction
exactly agrees with the Czachor's result.

To measure the degree of violation of the Bell's inequality, we
then evaluate the so-called Bell observable for the Bell states.
The degree of violation decreases under Lorentz boost. However,
this is the case when one evaluates the Bell observable with the
same spin measurement directions as in the non-relativistic lab
frame.
In fact, we show that the Bell's inequality is still
maximally violated in a Lorentz boosted frame, if we properly
choose new set of spin measurement directions. We show this
following the reasoning of Terashima and Ueda \cite{tu} for all
the Bell states.

In the non-relativistic case, maximal violation of the Bell's
inequality implies full entanglement of a given state. Thus we may
infer that the restoration of maximal violation of the Bell's
inequality in a Lorentz boosted frame indicates the preservation
of the entanglement information in a certain form even under
Lorentz boost.

We check this idea by investigating how the EPR correlation of the
spin singlet whose spins are up and down in the $z$-direction
changes under a ultra-relativistic Lorentz boost that reaches the
speed of light in the $x$-direction.
As we discussed in section IV, the new spin measurement directions
which give the maximal violation of the Bell's inequality, i.e.,
which preserve the expectation value of the joint spin
measurement, become the $+x$ and $-x$ directions when the original
directions for the joint spin measurement in the lab frame are
both in the $+z$ direction.
Namely, the perfect anti-correlation of the spin singlet becomes
the perfect correlation under a ultra-relativistic Lorentz boost
perpendicular to the original spin directions.

However, one can also check that for the unchanged Bell states
under Lorentz boost such as $| \Phi_{p}^{(-)} > ~, ~
|\Psi_{p}^{(+)} > $ in (\ref{ce47}), the form of correlation does
not change.

We thus conclude that the entanglement information is preserved as
a form of correlation information determined by the transformation
characteristic of the Bell state in use.

Finally, we would like to notice that even though the Lorentz
boost is not unitary, our relativistic spin observable (\ref{ce7})
in section II is determined by the rotation from the $z$-axis to
the direction of the particle's momentum $\vec{p}$. From this and
that the spin state is transformed by the Wigner rotation which is
unitary, we can infer that the restoration of the maximal
violation of Bell's inequality by adjusting the measurement axes
might be expected from the unitarity of the rotation. This is in a
sense similar to the cases of Refs. \cite{am,tu} where the
unitarity of the spin state transfomation given by the Wigner
rotation ensures the same result, since their spin observables are
not changed.
\\

\vspace{5mm}

\noindent
{\Large \bf Acknowledgments}

\vspace{5mm}

\noindent We would like to thank B.H. Lee for comment on our
calculational error in section II. This work was supported in part
by Korea
Research Foundation, Grant No. KRF-2002-042-C00010. \\


\end{document}